\documentclass[runningheads,a4paper]{llncs}
\usepackage{mathcmds}
\usepackage[table,pdftex]{xcolor}
\usepackage{subcaption}
\usepackage[colorlinks=true,citecolor=blue,urlcolor=black]{hyperref}
\usepackage[colorinlistoftodos]{todonotes}
\usepackage{tikz}
\usepackage{comment}
\usepackage{booktabs}
\usepackage[export]{adjustbox}
\usepackage{siunitx}
\usetikzlibrary{arrows.meta}

\begin{document}

\mainmatter  
\title{Improving RetinaNet for CT Lesion Detection with Dense Masks from Weak RECIST Labels}
\titlerunning{Improving RetinaNet for CT Lesion Detection}

\newif\ifreview

\ifreview
\author{Paper ID$:$ 338}
\institute{}
\else
\author{Martin Zlocha \and Qi Dou \and Ben Glocker}
\institute{Biomedical Image Analysis Group, Imperial College London, UK\\
\email{\{martin.zlocha15,qi.dou,b.glocker\}@imperial.ac.uk}}
\authorrunning{M.~Zlocha, Q.~Dou, and B.~Glocker}
\fi

\maketitle


\begin{abstract}
 Accurate, automated lesion detection in Computed Tomography (CT) is an important yet challenging task due to the large variation of lesion types, sizes, locations and appearances. Recent work on CT lesion detection employs two-stage region proposal based methods trained with centroid or bounding-box annotations. We propose a highly accurate and efficient one-stage lesion detector, by re-designing a RetinaNet to meet the particular challenges in medical imaging. Specifically, we optimize the anchor configurations using a differential evolution search algorithm. For training, we leverage the response evaluation criteria in solid tumors (RECIST) annotation which are measured in clinical routine. We incorporate dense masks from weak RECIST labels, obtained automatically using GrabCut, into the training objective, which in combination with other advancements yields new state-of-the-art performance. We evaluate our method on the public DeepLesion benchmark, consisting of 32,735 lesions across the body. Our one-stage detector achieves a sensitivity of 90.77\% at 4 false positives per image, significantly outperforming the best reported methods by over 5\%.
\end{abstract}

\section{Introduction}
\label{sec:introduction}

Detection and localization of abnormalities in Computed Tomography (CT) scans is a critical routine task for radiologists. Accurate, automated detection of suspicious regions could greatly support screening, diagnosis and monitoring of disease progression. Most previous work focuses on a specific type of lesion within a relatively constrained (anatomical) context, such as lymph nodes, lung nodules and brain microbleeds. Recently, Yan et al.~\cite{yan2018deep} pioneered the study of universal lesion detection and introduced today's largest data repository, i.e., the DeepLesion dataset.
Detecting diverse types of lesions across the body using one single model is very challenging due to the large variation of lesion types, sizes, locations and heterogeneous appearances. For example, DeepLesion consists of eight types of lesions with diameters ranging from 0.21 to 342.5 mm. In addition, the lesions may appear with limited contrast compared to nearby normal tissue, which further increases the difficulty of detecting subtle signs of disease.

Automated lesion detection has been central in medical image computing. Recent work employs two-stage methods with candidate proposal and false positive reduction steps. 
State-of-the-art performance on the DeepLesion benchmark has been achieved by Yan et al.~\cite{yan20183dce}.
They propose a two-stage, region-based method called 3DCE to effectively incorporate 3D context into 2D regional CNNs. Their method achieves a sensitivity of 85.65\% at 4 false positives per image, outperforming the popular detection method of Faster R-CNN~\cite{ren2015faster} on the same dataset.
However, their detection sensitivity for small lesions is much lower, which is an important limitation in the critical context of detecting early signs of diseases.

Some recent work take advantage of mask information for improving detection accuracy.
Jaeger et al.~\cite{jaeger2018retina} propose a Retina U-Net, showing that aggregating pixel-wise supervision to train the detector is helpful. Their method shows effectiveness in two scenarios, i.e., lung lesions in CT and breast lesions in MRI.
As pixel-wise annotations are tedious and expensive to obtain, Tang et al.~\cite{tang2019uldor} generate pseudo masks by fitting ellipses based on the response evaluation criteria in solid tumors (RECIST)~\cite{eisenhauer2009new} diameters.
Using a 2D Mask R-CNN~\cite{he2017mask} with generated lesion masks and other strategies,~\cite{tang2019uldor} achieves a sensitivity of 84.38\% at 4 false positives per image on DeepLesion dataset. Their pseudo-mask generation procedure relies heavily on the assumption of elliptical geometry of lesions, which may yield imprecise masks limiting the efficacy of dense supervision.

We propose a one-stage detector which directly localizes lesions without the need of candidate region proposals. To meet the specific challenge of detecting small lesions, we revisit the RetinaNet~\cite{lin2017focal} and optimize the feature pyramid scheme and anchor configuration by employing a differential evolution search algorithm. To enhance the model, we leverage high-quality dense masks obtained automatically from weak RECIST labels using GrabCut~\cite{rother2004grabcut}. Incorporating these generated masks into pixel-wise supervision shows great benefit for training the detector. In addition, we make use of the coherence between lesion mask predictions and bounding-box regressions to calibrate the detector outputs.
We further investigate recent strategies for boosting the detection performance, such as integrating attention mechanism into our feature pyramids. We evaluate the contributions of each part using the DeepLesion benchmark, achieving a new state-of-the-art sensitivity of 90.77\% at 4 false positives per image, significantly outperforming the currently best performing method 3DCE~\cite{yan20183dce} by over 5\%.

\section{Improving RetinaNet}
\label{sec:method}

\begin{figure}[t]
    \centering
    \includegraphics[width=0.95\linewidth]{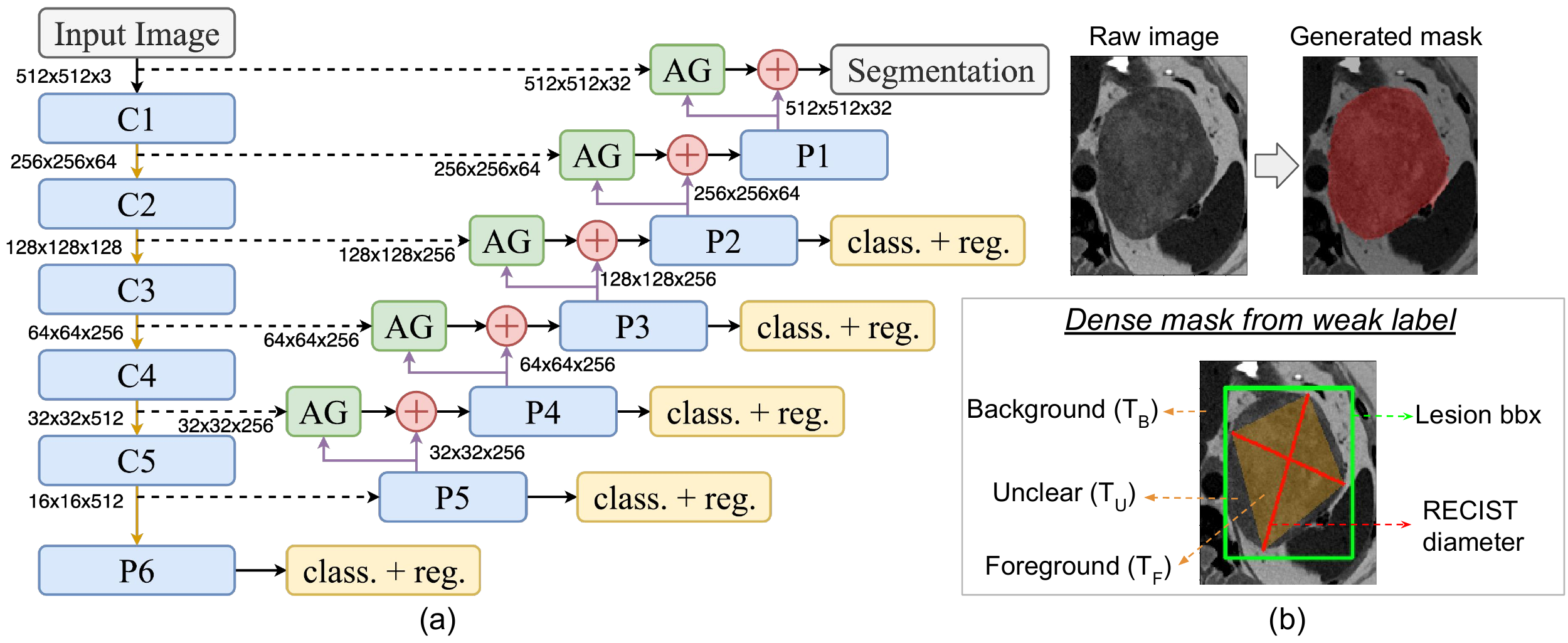}
    \vspace{-2mm}
    \caption{(a) Overview of our improved RetinaNet. (b) Automatic dense mask generation from weak RECIST diameters using GrabCut~\cite{rother2004grabcut}.}
    \vspace{-6mm}
    \label{fig:network}
\end{figure}

An overview of our proposed one-stage lesion detector is illustrated in Fig~\ref{fig:network} (a). We first describe the model design before elaborating on how we obtain dense masks from weak RECIST labels and incorporate them into training process. We then show the attention mechanism for further improving detection performance.

\vspace{-3mm}
\subsection{Model Design with Optimized Anchor Configuration}

The backbone of our approach is a RetinaNet~\cite{lin2017focal}, a recent one-stage method for object detection.
The use of a focal loss addresses the common problem of class imbalance in detection tasks.
The feature pyramids and lateral connections with a top-down architecture~\cite{lin2017feature} are adopted for detecting objects at different scales.
This is an important difference with methods such as 3DCE~\cite{yan20183dce}, since the feature pyramids can effectively capture information about lesions of varying sizes including very small ones.
Our specific network follows the structure of VGG-19~\cite{simonyan2014very}. 
We also explored ResNet-50 as used originally, but its performance was worse on DeepLesion, which is in line with results reported in \cite{yan2018deeplesion}. 

The anchor configuration is crucial for the detector, and we find the default anchor sizes (32, 64, 128, 256 and 512), aspect ratios (1:2, 1:1 and 2:1) and scales ($2^\frac{0}{3}$, $2^\frac{1}{3}$ and $2^\frac{2}{3}$) 
turn out to be ineffective for detecting lesions of small size and large ratios.
We employ a differential evolution search algorithm~\cite{storn1997differential} to optimize ratios and scales of anchors on the validation set. This algorithm iteratively improves a population of candidate solutions with regard to an objective function. New solutions are created by combining existing ones. We aim to find the best anchor settings for 3 scales and 5 ratios.
The objective is to maximise the overlap between the lesion bounding-box and the best anchor on the validation dataset.
We fix one ratio as 1:1, and define other ratios as reciprocal pairs (i.e., if one ratio is $1 \! : \!\gamma$ then another is $\gamma \! : \! 1$). Thus, we need to optimise only five variables, i.e, two ratio pairs and three scales.
When initialising the population of candidate solutions, all scales are bounded to a range of $[0.4, 1.6]$ and the two ratios are respectively bounded in $[1, 2]$ and $[2, 4]$.
We obtain optimal scales as 0.425, 0.540 and 0.680, and ratios of 3.27:1, 1.78:1, 1:1, 1:1.78, 1:3.27, which fits objects of small size and large ratios. Anchor sizes remain as (32, 64, 128, 256 and 512). These optimised configurations are then used for training the detector.

\vspace{-3mm}
\subsection{Dense Mask Supervision from Weak RECIST Labels}

Although annotations of bounding-boxes are relatively easy to obtain, there are other ``weak'' labels which are routinely generated in clinical practice, such as RECIST diameters. RECIST is used to track lesion growth, and consists of a pair of diameters to measure the lesion extent (cf. Fig.~\ref{fig:network}(b)). To leverage this highly valuable information, we automatically generate dense lesion masks from RECIST labels (provided in the DeepLesion dataset) using GrabCut~\cite{rother2004grabcut}.
We initialize a trimap into background ($T_B$), foreground ($T_F$) and unclear ($T_U$) pixels. A segmentation mask is generated based on iterative graph-cuts. Initialization can largely affect the final result, as it defines the Gaussian mixture models capturing the foreground and background intensity distributions.

Cai et al.~\cite{cai2018accurate} previously adopt GrabCut to initialise lesion masks of the RECIST-slice for the task of weakly-supervised lesion segmentation in 3D. Their $T_B$ is set as pixels outside the bounding-box defined by RECIST axes, and $T_F$ is obtained by dilation of the diameters.
Such an initialisation may be sub-optimal, specifically, for large lesions, where a considerable number of lesion pixels, which are quite certain to belong to foreground, are outside the dilation and omitted in $T_F$.
For small lesions, the dilation has the risk of hard-labelling background pixels into $T_F$, which cannot be corrected in the optimization.

To achieve a higher-quality masks using GrabCut, we propose a new strategy, as illustrated in Fig.~\ref{fig:network}(b).
We build a quadrilateral by consecutively connecting the four endpoints of the RECIST diameters.
A pixel is labelled as foreground if it falls inside the quadrilateral.
As most lesions show convex outlines, this is a simple yet reliable strategy. With the annotation of bounding-box provided in the dataset, the
pixels outside the box are hard-labelled as background $T_B$. All remaining pixels are assigned to $T_U$ and estimated through GrabCut.


To exploit these generated dense labels, we add two more upsampling layers (connecting to P2 and P1) and a segmentation prediction layer to the detector.
Skip connections are employed by fusing features obtained from C1 (via a  $1\times 1$ convolution) and input (via two $3\times 3$ convolutions), as shown in Fig.~\ref{fig:network}(a).
To retain sufficient resolution of feature maps for small lesions, we shift the sub-network operation (i.e., classification and regression) to pyramid levels of P2-P6 from P3-P7.
Using dense supervision to help detection task shares the idea with Retina U-Net~\cite{jaeger2018retina}, where we avoid the need for tedious labelling, as our dense masks are automatically generated from labels that are already recorded in clinical routine.
Additionally, we leverage the IoU between a bounding-box around the predicted segmentation mask and the directly regressed box (yellow sub-networks in Fig.~\ref{fig:network}), to calibrate the prediction probability $\tilde{p} = p \times (1+\text{IoU})$ of a lesion.
High coherence between segmentation and detection results indicates high confidence in lesion prediction, and benefits sensitivity at low FP rates.

\vspace{-3mm}
\subsection{Attention Mechanism for Gated Feature Fusion}


A recent attention gate (AG) model proposed by Schlemper et al.~\cite{schlemper2019attention} learns to focus on target structures by producing an attention map. According to this work, this may be beneficial for small, varying structures.
We explore AGs to filter feature responses propagated through skip connections and use features upsampled from coarser scale as the gating signal.
The AG module only uses $1\!\times\!1$ convolutions and produces a single attention map, which makes it computationally light-weight.
The output of AG is the element-wise multiplication of the attention map and the feature map from the skip connection. 


\vspace{-3mm}
\paragraph{\textbf{Training:}} We follow the loss used in original RetinaNet for detection, and our segmentation uses focal loss with cross-entropy.
We employ the Adam optimizer with a learning rate of $10^{-4}$ which is reduced during training by a factor of 10 when the mean average precision (mAP) has not improved for 2 consecutive epochs. The batch size is 4 during training. To reduce overfitting, early stopping is used if the mAP has not improved for 4 consecutive epochs on the validation set. We use an NVIDIA GeForce GTX 1080 for training and testing.

\vspace{-2mm}
\section{Experiments}
\label{sec:experiments}

\subsection{Dataset, Pre-Processing, and Augmentation}
The public DeepLesion dataset~\cite{yan2018deep} consists of 32,120 axial CT slices from 10,594 studies of 4,427 unique patients. There are $1 \! \sim \! 3$ lesions in each slice, adding up to 32,735 lesions altogether. 
For each lesion, there is usually 30mm of extra slices above and below the key slice to provide contextual information.
In most cases, the slices have 1 or 5 mm thickness, but this varies with some being 0.625 or 2 mm.
The 2D bounding-boxes and RECIST diameters for lesions are annotated on the key slice.
The dataset covers a wide range of lesions from lung, liver, mediastinum (mainly lymph nodes), kidney, pelvis, bone, abdomen and soft tissue.
Sizes vary significantly with diameters ranging from 0.21 to 342.5 mm.

We perform lightweight pre-processing where images are resized into $512\!\times\!512$ pixels, resulting in a voxel-spacing between 0.175 and 0.977 mm with a mean of 0.802 mm. The Hounsfield units (HU) are clipped into the range of $[-1024, 1050]$. We normalize the intensities to the range of $[-1,1]$ as input to the network.
In our experiments, we use three adjacent slices after resampling to 2 mm thickness. In rare cases where the neighboring slice of the lesion slice is not provided, we duplicate the lesion slice to fill the missing input channels. 
We use data augmentation where images are flipped in horizontal and vertical directions with 50\% chance. We also use random affine transformations with rotation/shearing up to 0.1 radians, and scaling/translation up to 10\% of the image size.



\vspace{-3mm}
\subsection{Detection Results on DeepLesion Benchmark}

The DeepLesion dataset is provided with splits into 70\% for training, 15\% for validation, and 15\% for testing. Thus, our results can be directly compared with numbers reported in the literature. 
The current best results have been achieved by Yan et al.~\cite{yan20183dce} and  Tang et al.~\cite{tang2019uldor}. We also quote their provided baseline performance using popular detection methods, i.e., Faster R-CNN~\cite{ren2015faster} (reported in~\cite{yan20183dce}) and Mask R-CNN~\cite{he2017mask} (reported in~\cite{tang2019uldor}). We further provide the results of our own baseline RetinaNet~\cite{lin2017focal} using its default configuration.
A predicted box is regarded as correct if its IoU
with a ground truth box is larger than $0.5$. 

\begin{figure}[t]
  \centering
  \includegraphics[width=1\linewidth]{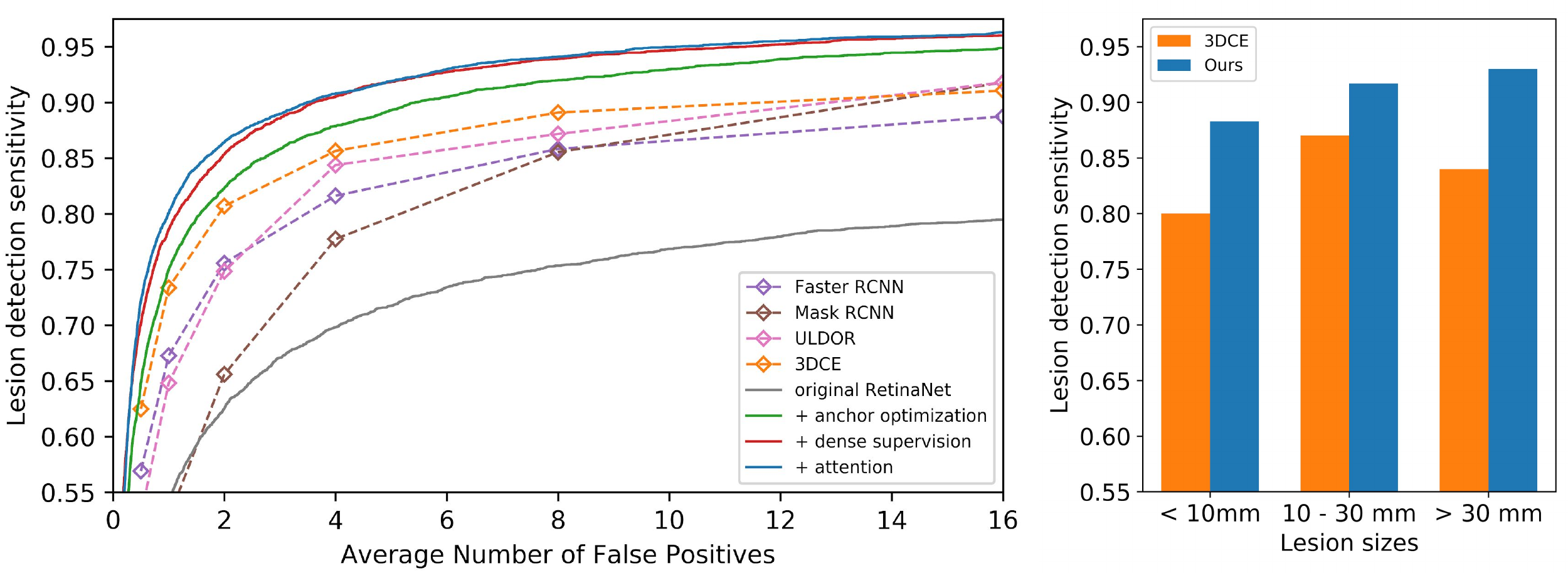}
  \vspace{-4mm}
  \caption{\textbf{Left}: FROC curves for our improved RetinaNet variants and baselines on DeepLesion dataset. \textbf{Right}: Per lesion size results compared to 3DCE~\cite{yan20183dce}.}
  \vspace{-3mm}
  \label{fig:froc}
\end{figure}

In Table~\ref{tab:detection}, we present the lesion detection sensitivities at different false positives (FP) per image. Our improved RetinaNet consistently outperforms existing methods across all FP rates. Specifically, sensitivity at 4 FPs, which is commonly reported in the literature, we achieve a sensitivity of 90.77\%, which is a 5.12\% improvement over 3DCE~\cite{yan20183dce} and 6.39\% over ULDor~\cite{tang2019uldor}.
The free-response receiver operating characteristics (FROC) curves of different methods are shown on the left in Fig.~\ref{fig:froc}. We observe that optimized networks for lesion detection are generally better than out-of-the-box detectors, such as Faster R-CNN, Mask R-CNN and RetinaNet.
When comparing sensitivity at low FP rates, our improved models perform much better than others, indicating the benefit of task-specific optimization and incorporation of additional mask information.

The sensitivity for detecting different sizes of lesions at 4 FPs are shown on the right in Fig.~\ref{fig:froc}. We divide the lesions into three size groups according to the diameter, following~\cite{yan20183dce} for direct comparison. For small lesions with diameters less than 10 mm, our sensitivity is 88.35\% compared to 80\% for 3DCE.
Using a feature pyramid to retain responses from small lesions together with dense supervision with focal loss seems beneficial for detecting subtle signs of disease.
While 3DCE uses richer 3D context, this seems less helpful for small, local structures.
Our model works well across all lesion sizes where we improve sensitivity from 87\% to 91.73\% for lesions of $10 \! \sim \! 30$ mm, and from 84\% to 93.02\% for lesions larger than 30 mm when compared to 3DCE.

We also record average inference time for each image during testing, as listed in Table~\ref{tab:detection}.
All the detection results are obtained using a single network without model ensemble nor test augmentation.
Our one-stage detector is highly efficient eliminating the need of generating lesion proposals. The integration of dense supervision and attention mechanism has minimal computational overhead, taking about 41 ms for each image.
Runtimes are reported for 3DCE and Faster R-CNN in \cite{yan20183dce}, but a comparison is only indicative due to different GPUs being used.

\begin{table}[t]
    \centering
    \caption{Detection performance of different methods and our ablation study.}
    \vspace{1mm}
    \begin{tabular}{c|p{1.0cm}<{\centering}p{1.0cm}<{\centering}p{1.0cm}<{\centering}p{1.0cm}<{\centering}p{1.0cm}<{\centering}p{1.0cm}<{\centering}|p{1.3cm}<{\centering}}
    \toprule
     Methods                       & 0.5 & 1 & 2 & 4 & 8 & 16 & runtime \\
     \midrule
     Faster R-CNN~\cite{ren2015faster}      & 56.90 & 67.26 & 75.57 & 81.62 & 85.83 & 88.74 & 32 ms\\
     Mask R-CNN~\cite{he2017mask}              & 39.82 & 52.66 & 65.58 & 77.73 & 85.54 & 91.80 & - \\
     ULDor (Tang et al.~\cite{tang2019uldor})  & 52.86 & 64.80 & 74.84 & 84.38 & 87.17 & 91.80 & -\\
     3DCE (Yan et al.~\cite{yan20183dce})      & 62.48 & 73.37 & 80.70 & 85.65 & 89.09 & 91.06 & 114 ms\\
     \midrule
     original RetinaNet~\cite{lin2017focal}    & 45.80 & 54.17 & 62.50 & 69.80 & 75.34 & 79.48 & 28 ms \\
     + anchor optimization                     & 64.82 & 74.98 & 82.29 & 87.87 & 92.20 & 94.90 & 31 ms\\
     + dense supervision                       & 70.24 & 78.28 & 85.10 & 90.39 & 93.81 & 96.01 & 39 ms \\
     + attention gate                              & \textbf{72.15} & \textbf{80.07} & \textbf{86.40} & \textbf{90.77} & \textbf{94.09} & \textbf{96.32} & 41 ms \\
     \bottomrule
    \end{tabular}
    \vspace{-3mm}
    \label{tab:detection}
\end{table}

\begin{figure}[h]
\centering
\setlength{\tabcolsep}{1pt}
\def\arraystretch{0}
\begin{tabular}{cccc}
\\[-\ht\strutbox]
\subcaptionbox*{}{\includegraphics[width = 0.25\linewidth]{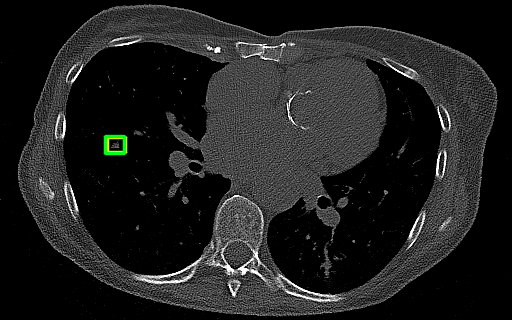}} &
\subcaptionbox*{}{\includegraphics[width = 0.25\linewidth]{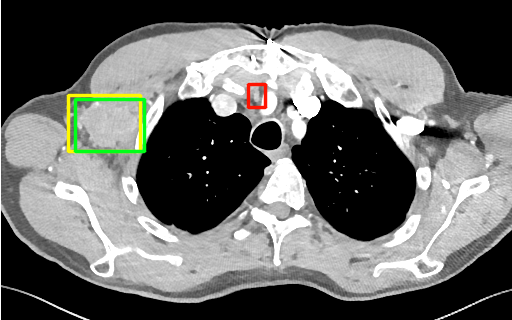}} &
\subcaptionbox*{}{\includegraphics[width = 0.25\linewidth]{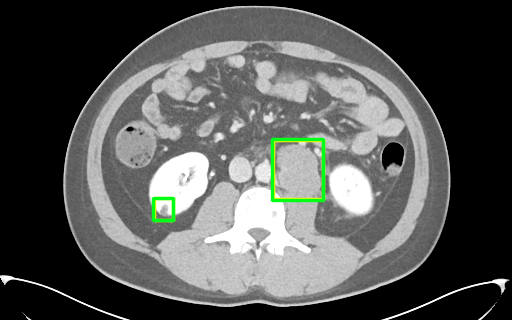}} &
\subcaptionbox*{}{\includegraphics[width = 0.25\linewidth]{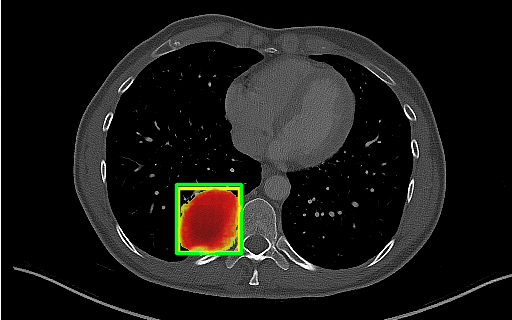}}
\\[-\ht\strutbox] \\[-0.8em]
\subcaptionbox*{}{\includegraphics[width = 0.25\linewidth]{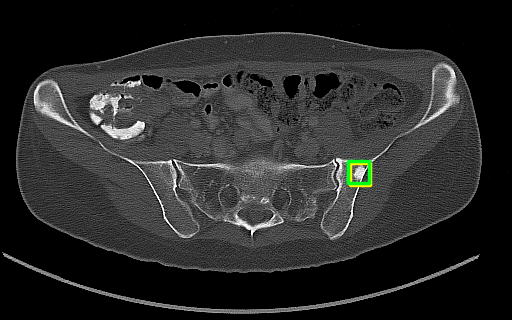}} &
\subcaptionbox*{}{\includegraphics[width = 0.25\linewidth]{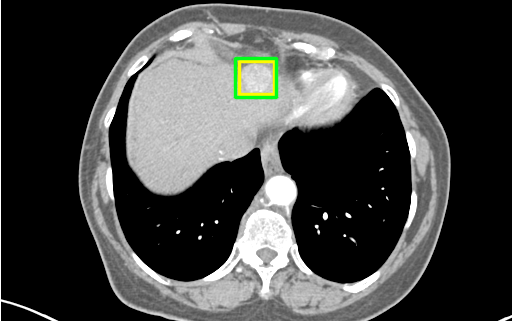}} &
\subcaptionbox*{}{\includegraphics[width = 0.25\linewidth]{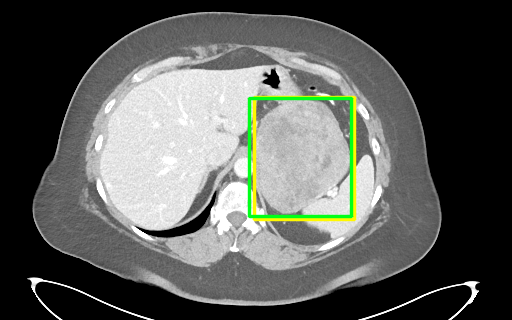}} &
\subcaptionbox*{}{\includegraphics[width = 0.25\linewidth]{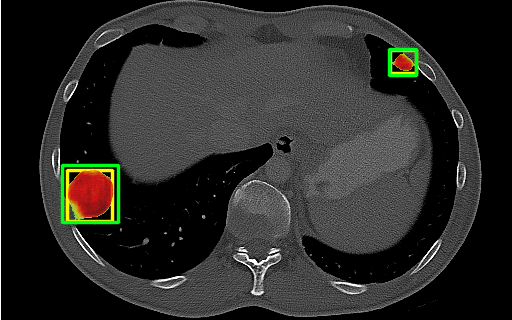}}
\\[-\ht\strutbox] \\[-0.8em]
\subcaptionbox*{}{\includegraphics[width = 0.25\linewidth]{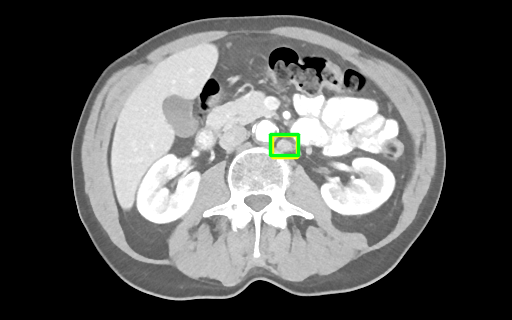}} &
\subcaptionbox*{}{\includegraphics[width = 0.25\linewidth]{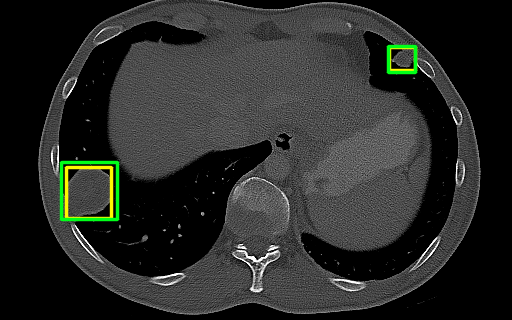}} &
\subcaptionbox*{}{\includegraphics[width = 0.25\linewidth]{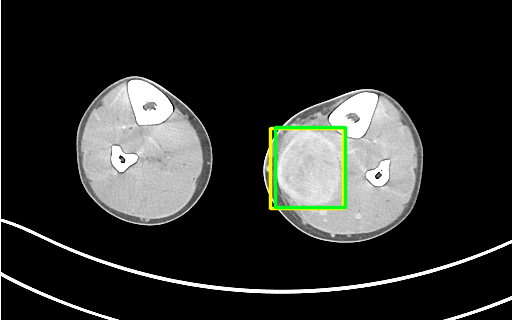}} &
\subcaptionbox*{}{\includegraphics[width = 0.25\linewidth]{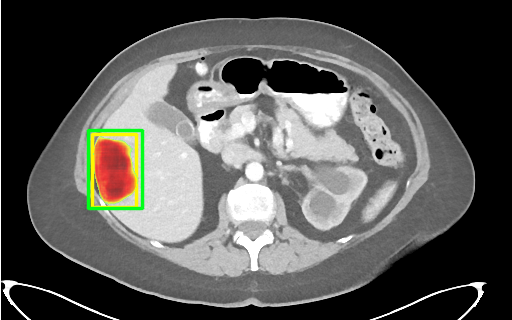}}
\\[-\ht\strutbox] \\[-0.8em]
\subcaptionbox*{}{\includegraphics[width = 0.25\linewidth]{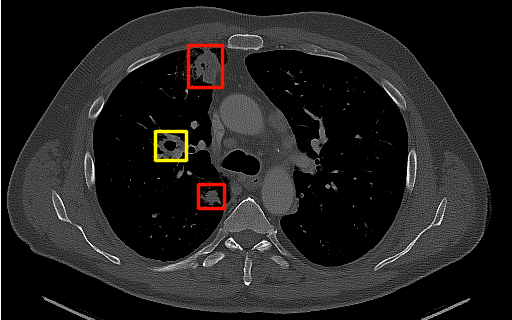}} &
\subcaptionbox*{}{\includegraphics[width = 0.25\linewidth]{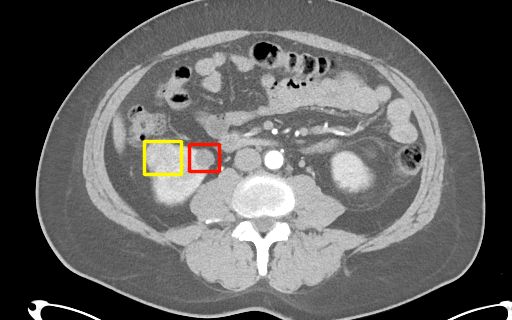}} &
\subcaptionbox*{}{\includegraphics[width = 0.25\linewidth]{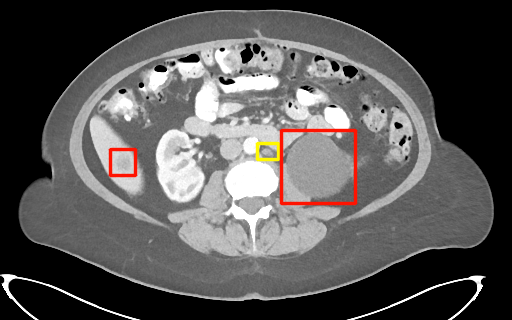}} &
\subcaptionbox*{}{\includegraphics[width = 0.25\linewidth]{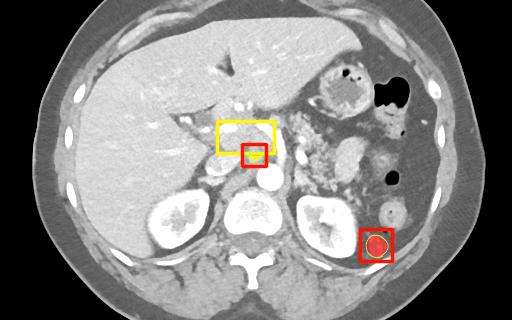}}
\end{tabular}
\vspace{-8mm}
\caption{Visual results for lesion detection at 0.5 FP rate using our improved RetinaNet. The first three columns show different sizes from small to large. The right column shows heatmaps from the segmentation layer overlaid on detections. Yellow boxes are ground truth, green are true positives, red are false positives. Last row shows intriguing failure cases with possibly incorrect ground truth.}
\label{fig:visual_results}
\vspace{-5mm}
\end{figure}

\vspace{-3mm}
\subsection{Contribution of Individual Improvements}

We investigate the individual impact of the proposed additions leading to our final improved RetinaNet. In an ablation study, we first evaluate the original RetinaNet~\cite{lin2017focal} with default settings, then incrementally add our improvements, i.e., automatic anchor optimization, dense supervision using lesion masks from weak labels, and attention mechanism.
Table~\ref{tab:detection} and Fig.~\ref{fig:froc} (left) summarizes these results. The original RetinaNet with default anchor configuration is performing poorly on the lesion detection task, indicating that out-of-the-box approaches from computer vision are sub-optimal. 
Remarkably, after employing the automatic search algorithm to optimize the anchor configuration, the simple RetinaNet already outperforms previous state-of-the-art. The sensitivity at 0.5 FP is 2.34\% higher than 3DCE and 11.96\% higher than ULDor.

Adding dense supervision with segmentation masks generated from RECIST diameters significantly boosts detection sensitivity across all FP rates, with 5.42\% improvement at 0.5 FP. The pixel-wise supervision adds an important training signal, providing more precise localization information in addition to bounding-boxes. Consistency between bounding-box regression and dense classification helps to reduce false positives. Finally, adding an attention mechanism further improves the performance, achieving a sensitivity of 90.77\% at 4 FPs, with an improvement of almost 10\% at 0.5 FP over the best reported results.

\begin{figure}[h]
\centering
\setlength{\tabcolsep}{1pt}
\def\arraystretch{0}
\begin{tabular}{cccc}
\\[-\ht\strutbox]
\subcaptionbox*{}{\includegraphics[width = 0.25\linewidth]{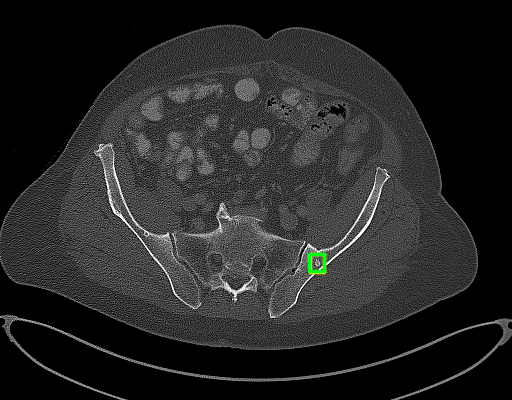}} &
\subcaptionbox*{}{\includegraphics[width = 0.25\linewidth]{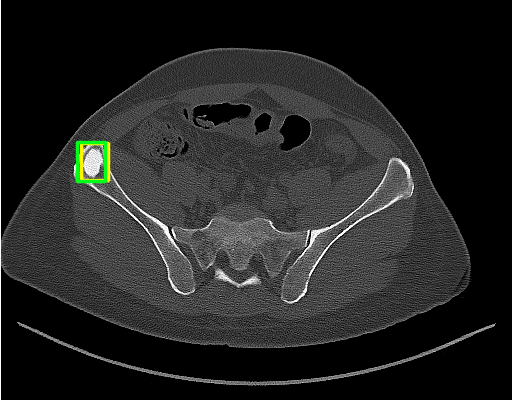}} &
\subcaptionbox*{}{\includegraphics[width = 0.25\linewidth]{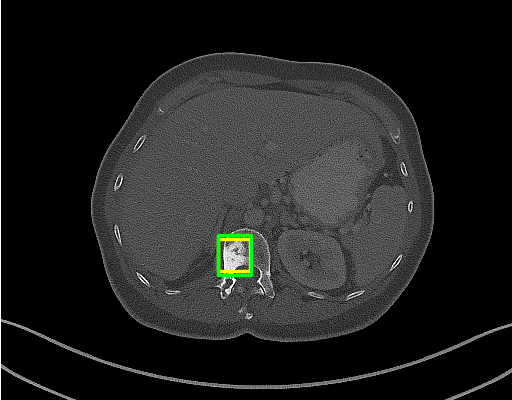}} &
\subcaptionbox*{}{\includegraphics[width = 0.25\linewidth]{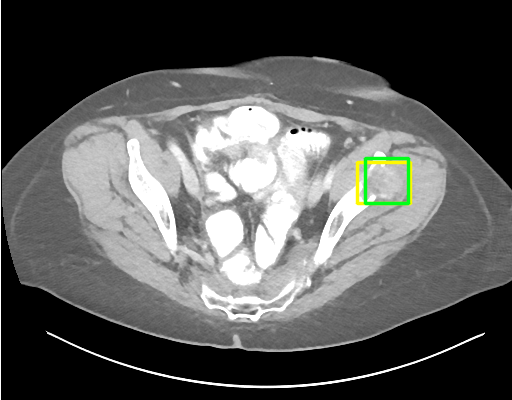}}
\\[-\ht\strutbox] \\[-0.8em]
\subcaptionbox*{}{\includegraphics[width = 0.25\linewidth]{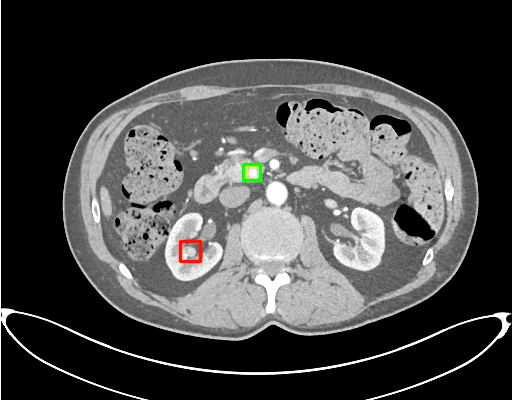}} &
\subcaptionbox*{}{\includegraphics[width = 0.25\linewidth]{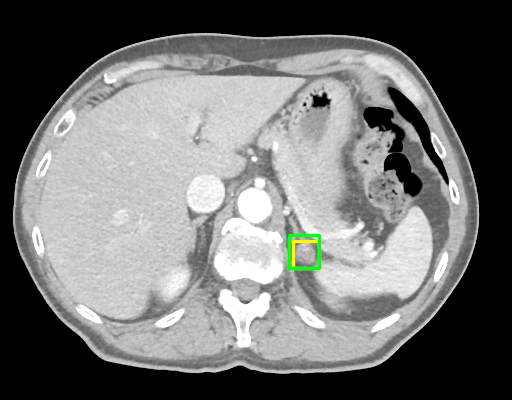}} &
\subcaptionbox*{}{\includegraphics[width = 0.25\linewidth]{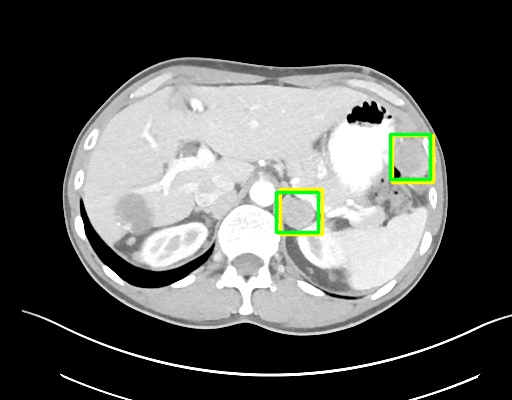}} &
\subcaptionbox*{}{\includegraphics[width = 0.25\linewidth]{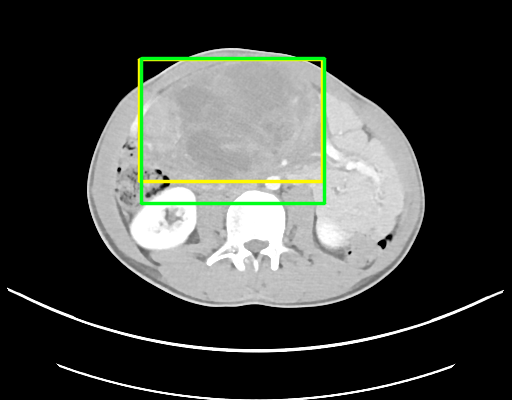}}
\\[-\ht\strutbox] \\[-0.8em]
\subcaptionbox*{}{\includegraphics[width = 0.25\linewidth]{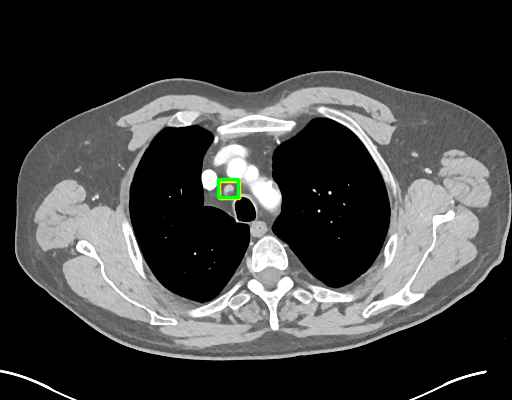}} &
\subcaptionbox*{}{\includegraphics[width = 0.25\linewidth]{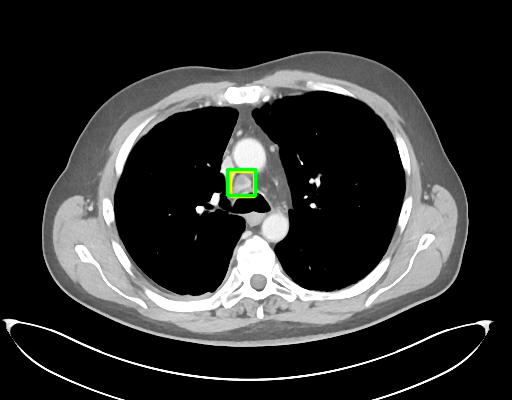}} &
\subcaptionbox*{}{\includegraphics[width = 0.25\linewidth]{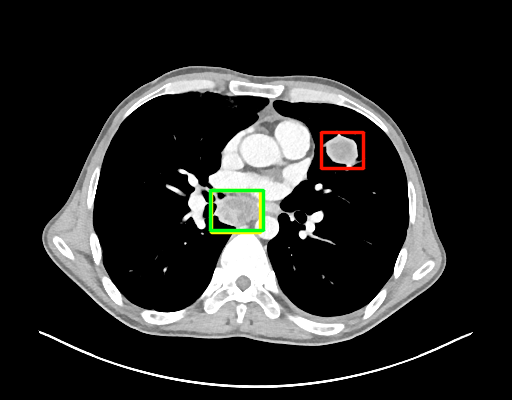}} &
\subcaptionbox*{}{\includegraphics[width = 0.25\linewidth]{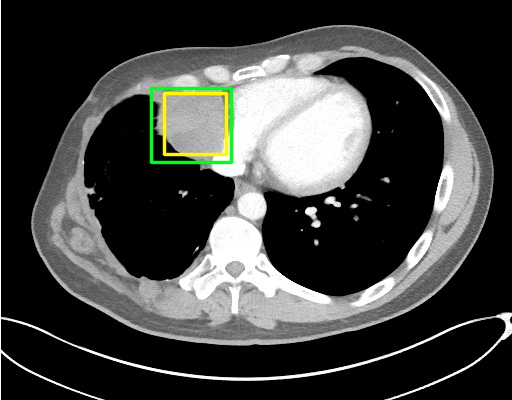}}
\\[-\ht\strutbox] \\[-0.8em]
\subcaptionbox*{}{\includegraphics[width = 0.25\linewidth]{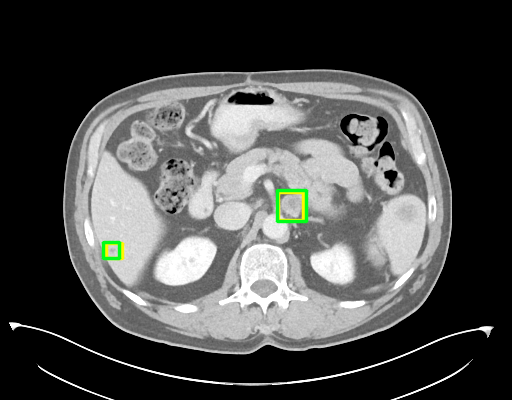}} &
\subcaptionbox*{}{\includegraphics[width = 0.25\linewidth]{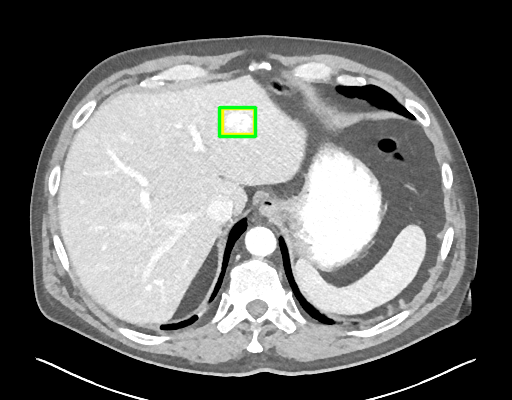}} &
\subcaptionbox*{}{\includegraphics[width = 0.25\linewidth]{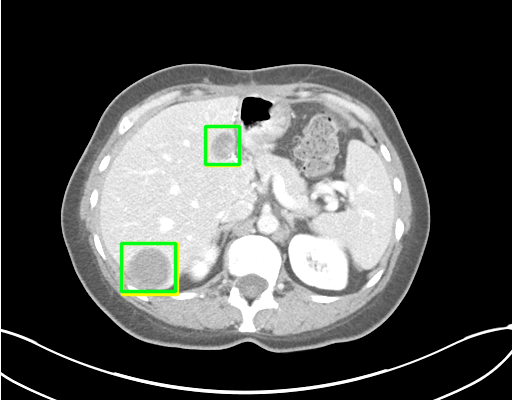}} &
\subcaptionbox*{}{\includegraphics[width = 0.25\linewidth]{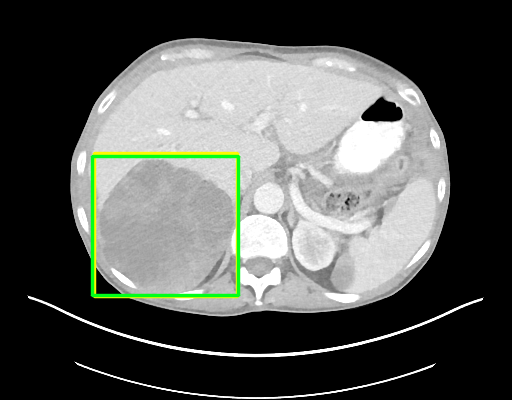}}
\\[-\ht\strutbox] \\[-0.8em]
\subcaptionbox*{}{\includegraphics[width = 0.25\linewidth]{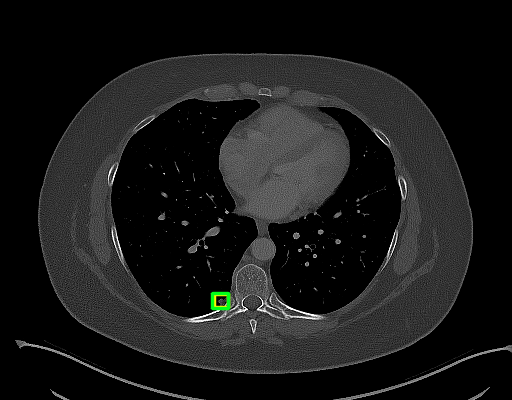}} &
\subcaptionbox*{}{\includegraphics[width = 0.25\linewidth]{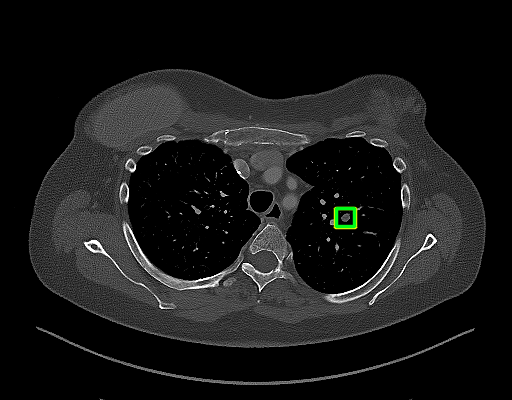}} &
\subcaptionbox*{}{\includegraphics[width = 0.25\linewidth]{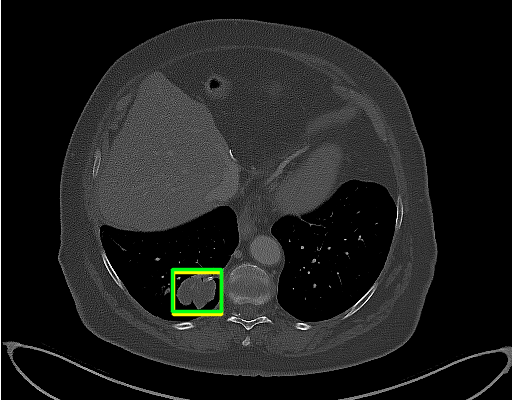}} &
\subcaptionbox*{}{\includegraphics[width = 0.25\linewidth]{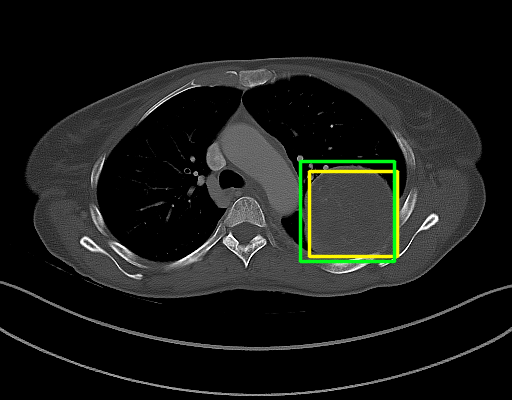}}
\\[-\ht\strutbox] \\[-0.8em]
\subcaptionbox*{}{\includegraphics[width = 0.25\linewidth]{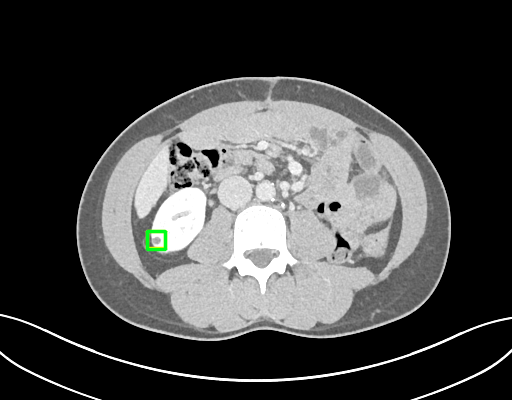}} &
\subcaptionbox*{}{\includegraphics[width = 0.25\linewidth]{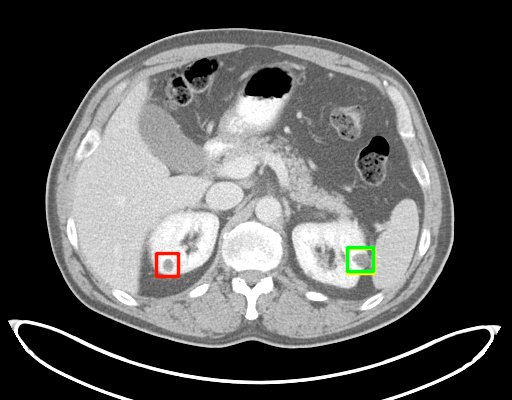}} &
\subcaptionbox*{}{\includegraphics[width = 0.25\linewidth]{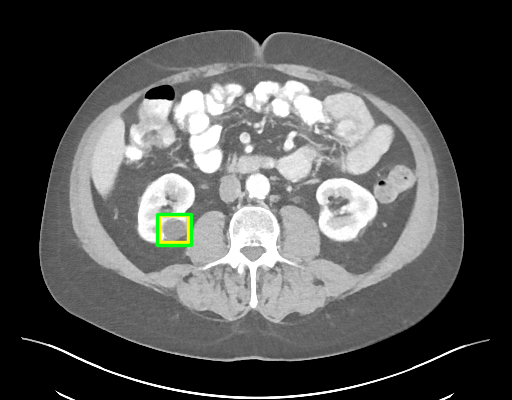}} &
\subcaptionbox*{}{\includegraphics[width = 0.25\linewidth]{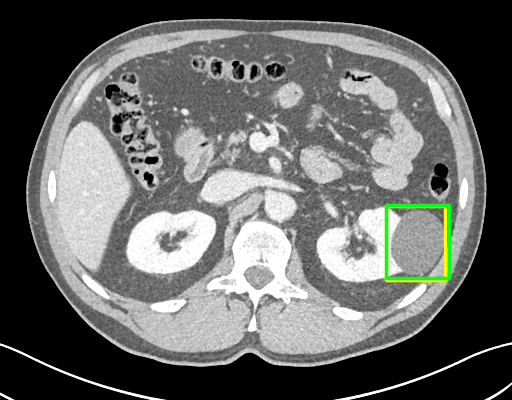}}
\\[-\ht\strutbox] \\[-0.8em]
\subcaptionbox*{}{\includegraphics[width = 0.25\linewidth]{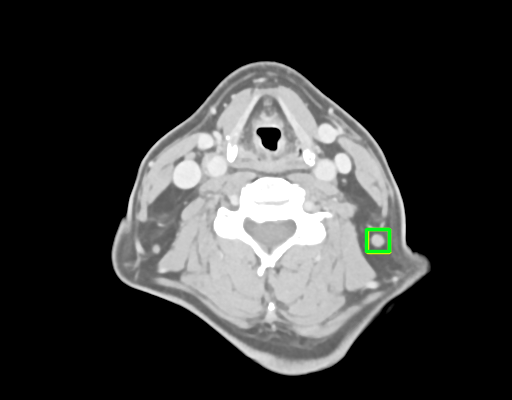}} &
\subcaptionbox*{}{\includegraphics[width = 0.25\linewidth]{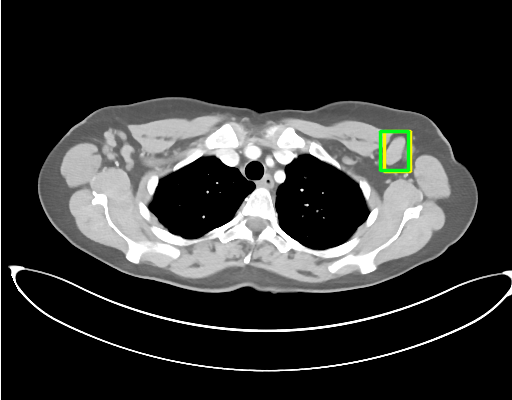}} &
\subcaptionbox*{}{\includegraphics[width = 0.25\linewidth]{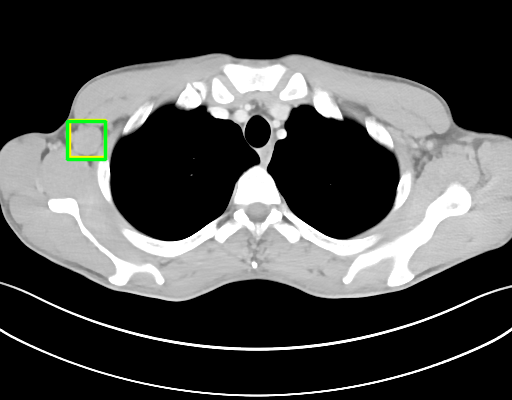}} &
\subcaptionbox*{}{\includegraphics[width = 0.25\linewidth]{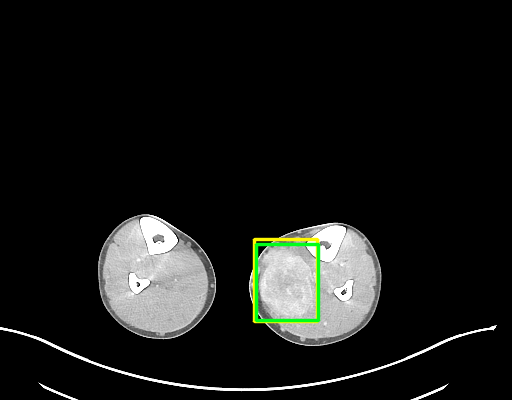}}
\\[-\ht\strutbox] \\[-0.8em]
\subcaptionbox*{}{\includegraphics[width = 0.25\linewidth]{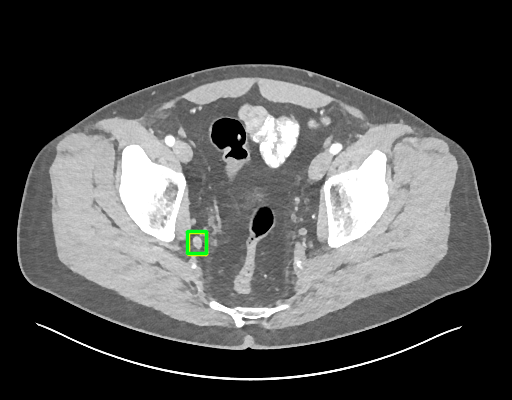}} &
\subcaptionbox*{}{\includegraphics[width = 0.25\linewidth]{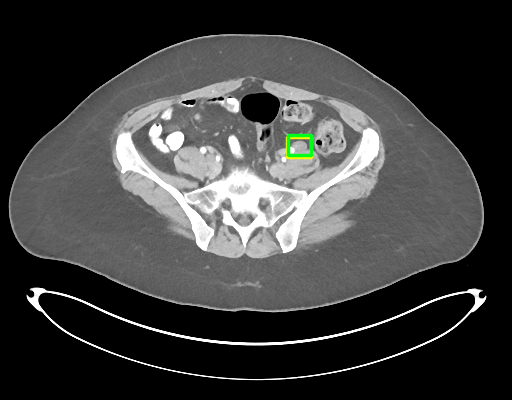}} &
\subcaptionbox*{}{\includegraphics[width = 0.25\linewidth]{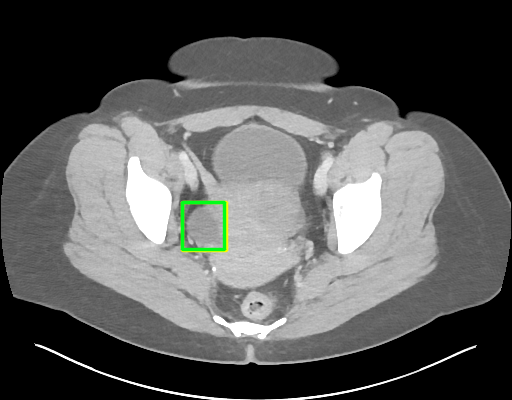}} &
\subcaptionbox*{}{\includegraphics[width = 0.25\linewidth]{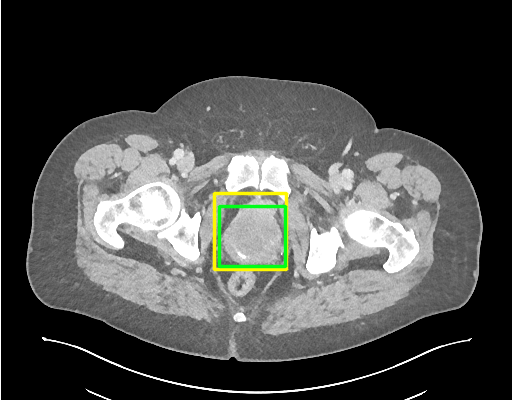}}
\\[-\ht\strutbox] \\[-0.8em]
\end{tabular}

\caption{More visual results for lesion detection at 0.5 FP rate using our improved RetinaNet. The rows correspond to bone, abdomen, mediastinum, liver, lung, kidney, soft tissue, and pelvis lesions,  respectively. Each row contains examples of lesions of different sizes ordered from smallest to largest. Yellow boxes are ground truth, green are true positives, red are false positives.}
\label{fig:visual_results2}
\vspace{-5mm}
\end{figure}

Visual examples of detected lesions on test images are shown in Figs.~\ref{fig:visual_results} and \ref{fig:visual_results2}.  Probability threshold is set to 0.3 yielding 0.5 FP per image. Lesions of various size, appearance and type are localized accurately. Segmentation masks look sensible, indicating good quality of the automatically generated dense labels for training.

\section{Conclusion}
\label{sec:conclusion}

Our improved RetinaNet shows impressive performance on CT lesion detection outperforming state-of-the-art by a significant margin. Interestingly, we could show that by task-specific optimization of an out-of-the-box detector we already achieve results superior than the best reported in the literature. Exploitation of clinically available RECIST annotations bears great promise as large amounts of such training data should be available in many hospitals. With a sensitivity of about 91\% at 4 FPs per image, our system may reach clinical readiness. Future work will focus on new applications such as whole-body MRI in oncology.

\section*{Acknowledgements}
This project has received funding from the European Research Council (ERC) under the European Union's Horizon 2020 research and innovation programme (grant agreement No 757173, project MIRA, ERC-2017-STG).
\bibliographystyle{splncs04_trunc}
\bibliography{cites_optimized}

\end{document}